\documentclass{aastex}
\usepackage{spr-astr-addons}
\usepackage{url}

\RequirePackage{color}
\renewcommand{\vec}[1]{\mbox{\boldmath $#1$}}

\newcommand{\emaila}{wilhelm@mps.mpg.de}
\newcommand{\emailb}{bnd.app@iitbhu.ac.in}
\renewcommand{\vec}[1]{\mbox{\boldmath $#1$}}
\def\arcsec{\hbox{$^{\prime\prime}$}}

\newcommand*\Del{\mathrm{\Delta}}                 
\newcommand{\rmd}{ {\ \mathrm d} }
\newcommand{\uvec}[1]{\hat{\vec #1}}
\newcommand{\m}{ {\ \mathrm m} }
\newcommand{\s}{ {\ \mathrm s} }
\newcommand{\kg}{ {\ \mathrm {kg}} }

\newcommand{\grad}{{\bf \nabla } }
\begin{document}

\title{Gravitational redshift and the vacuum index of refraction}
\shorttitle{Gravitational redshift}
\shortauthors{K. Wilhelm, B.N. Dwivedi}

\author{Klaus Wilhelm}
\affil{Max-Planck-Institut f\"ur Son\-nen\-sy\-stem\-for\-schung
(MPS),\\ Justus-von-Liebig-Weg 3, 37077 G\"ottingen, Germany\\ \emaila}

\and

\author{Bhola N. Dwivedi}
\affil{Department of Physics, Indian Institute of Technology
(Banaras Hindu University), Varanasi-221005, India\\ \emailb}

Last updated on \today

\vspace{1cm}

\begin{abstract}
A physical process of the gravitational redshift was described in an earlier
paper \citep{WilDwi14} that did not require any information for the emitting
atom neither on the local gravitational potential~$U$ nor on the speed of
light~$c$. Although it could be shown that the correct energy shift of the
emitted photon resulted from energy and momentum conservation principles and
the speed of light at the emission site, it was not obvious how this speed is
controlled by the gravitational potential. The aim of this paper is to
describe a physical process that can accomplish this control. We determine the
local speed of light~$c$ by deducing a gravitational index of
refraction~$n_{\rm G}$ as a function of the potential~$U$ assuming a specific
aether model, in which photons propagate as solitons. Even though an atom
cannot locally sense the gravitational potential~$U$
\citep[cf.][]{Mueetal}, the gravitational redshift will nevertheless be
determined by~$U$ \citep[cf.][]{Woletal}\,---\,mediated by the local speed of
light~$c$.
\end{abstract}

\keywords{Gravitation, impact model, aether, gravitational index of
refraction}

\section{Introduction} 
\label{s:introd}
\label{sec:intro}

The study of the gravitational redshift, predicted for solar radiation by
\citet{Ein08}, is still an important subject in modern physics and
astrophysics \citep[e.g.,][]{Kol04,Neg05,Lae09,Choetal,Pasetal,Tur13}.
The displacement of metallic lines to the violet observed in the laboratory
in comparison with the corresponding solar lines had first been noted by
\citet{Row96} and \citet{Jew96} \citep[cf.][]{Hen93a}.
Measurements of the small gravitational redshift of solar
spectral lines are inherently difficult, because many processes in the
atmosphere of the Sun can influence the spectrum. In particular, the high
speeds of the emitting plasmas lead to line shifts due to the classical Doppler
effect \citep[cf.][]{Hen93b}.
Nevertheless, early observations confirmed Einstein's prediction in general
\citep{StJ28,BlaRod,Bra62,Bra63,Sni70} \citep[cf.][]{Hen96}.
Improved observational techniques \citep[e.g.,][]{LoPetal,Cacetal,TakUen},
have established a shift of solar lines of
%
\begin{eqnarray}
c_0\,\frac{\Del \lambda}{\lambda} \approx 600\m\s^{-1} ~,
\label{shift_600}
\end{eqnarray}
where $c_0 = 299\,792\,458\m\s^{-1}$ is the speed of light in vacuum
remote from any masses and $\lambda$ the wavelength of the electromagnetic
radiation.

The gravitational potential~$U$ at a distance~$r$ from a spherical body with
mass~$M$ is constraint in the weak-field approximation for non-relativistic
cases \citep[cf.][]{LanLif} by
%
\begin{eqnarray}
- 1  \ll \frac{U}{c^2_0} = - \frac{G_{\rm N}\,M}{c^2_0~r} \le 0 ~,
\label{potential}
\end{eqnarray}
where $G_{\rm N} = 6.67554(16) \times 10^{-11}\kg^{-1}\m^3\s^{-2}$
is Newton's constant of gravity \citep{Quietal}.
A definition of a reference potential in line with
Eq.~(\ref{potential}) is $U_0 = 0$ for $r = \infty$.

In an attempt to describe the physical process(es) that lead to the
gravitational redshift,
\citet{Woletal} and \citet{Mueetal} disagreed on whether the frequency
of an atomic clock is sensitive to the
gravitational potential~$U$ (according to Wolf et al.) or, as suggested by
M\"uller et al., to the local gravity field $\vec{g} = \grad U$.
Support for the first alternative can be found in many publications
\citep[e.g.,][]{Ein08,Lau20,Sch60,Wil74,Okuetal,SinSam}, but it is, indeed,
not obvious how an atom can locally sense the gravitational potential~$U$.

Experiments on Earth~\citep{PouReb,Craetal,KraLue,PouSni},
in space~\citep{Vesetal,BauWey} and in the Sun-Earth
system~\citep{StJ28,BlaRod,Bra62,Bra63,Sni72,LoP91,Cacetal,TakUen} have,
however, quantitatively confirmed in the static weak field approximation a
relative frequency
shift of
%
\begin{eqnarray}
\frac{\nu - \nu_0}{\nu_0} = \frac{\Del \nu}{\nu_0}
\approx \frac{\Del U}{c^2_0} = \frac{U - U_0}{c^2_0}~,
\label{shift}
\end{eqnarray}
where $\nu_0 = c_0/\lambda_0$ is the frequency of the radiation emitted by a
certain transition at~$U_0$ and $\nu$ the observed frequency there, if the
emission caused by the same transition had occurred at a potential~$U$.

In addition to the redshift, the deflection of light near gravitational
centres is of fundamental importance. For a close solar fly-by
\citet{Sol04} and
\citet{Ein11} obtained $0.87\arcsec$ under the assumption
that radiation would be affected in the same way as matter.
\emph{Twice} this value was then derived in the framework of the General
Theory of Relativity \citep[GTR,][]{Ein16}\footnote{It is of interest in
the context of this paper that Einstein employed Huygens' Principle in his
calculation of the deflection.},
and later by \citet{Sch60} using the equivalence principle and STR.
The high value was confirmed
during the total solar eclipse in 1919 for the first time \citep[]{Dysetal}.
This and later observations have been summarized by \citet{Mik59} and
combined to a mean value of
approximately 2\arcsec.

\section{Graviton interactions} 
\label{sec:graviton}

A model of gravitational interactions based on a modified impact concept
has been proposed for massive bodies \citep[][Paper~1]{WiWiDw},
and the difficulties of the old theory proposed by
Nicolas \citet{Fat90} \citep[cf.][]{Bop29,Gag49} have been
considered in the light of the Special Theory of Relativity
\citep[STR,][]{Ein05a} and the non-local behaviour of virtual
particles \citep[cf.][]{NimSta}. The basic idea is that impacting
\emph{gravitons}\,---\,originally named \emph{quadrupoles}\,---\,with no mass
and a speed $c_0$ are absorbed by massive particles and re-emitted with
reduced energy~$T^-_{\rm G}$ according to
%
\begin{eqnarray}
T^-_{\rm G} = T_{\rm G}\,(1 - Y) ~,
\label{Eq:reduced_energy}
\end{eqnarray}
where $T_{\rm G}$ is the energy of a graviton in the background flux and
$0 < Y \ll 1$.
A spherically symmetric emission of a liberated graviton with a reduction
parameter~$Y$ had been assumed in Paper~1. Further studies
have, however, indicated that an anti-parallel emission with respect to the
incoming graviton is more appropriate, because conflicts with the energy and
momentum conversation principles in closed systems can be avoided by the
second choice.
Newton's law of gravitation could be explained with this model,
however, a secular mass increase of matter was a consequence
of its application. This poses the question of how the interaction of gravity
with photons can be understood, since the photon mass is in all likelihood
zero.\footnote{A zero mass of photons follows from the STR and a speed of light
in vacuum~$c_0$ constant for all frequencies. \citet{Ein05b} used
,,Lichtquant'' for a quantum of electromagnetic radiation;
the term ``photon'' was introduced by \citet{Lew26}.
With various methods the photon mass could be constrained to
$m_\nu < 10^{-49}~\kg$ \citep{GolNie,Amsetal}.}
If the mass of a photon is indeed zero, the interaction process must be
different.
An initial attempt at solving that problem has been made in Paper~2
\citep{WilDwi13} and is summarized here under the assumption of an
anti-parallel re-emission, both for massive particles and photons.

A physical process will then be outlined that provides information on
the gravitational potential~$U$ at the site of a photon emission. This aspect
had not been covered in our earlier paper on the gravitational redshift
\citep{WilDwi14}.

Interactions between massive bodies have been treated in Paper~1 with an
absorption rate of \emph{half} the intrinsic de Broglie frequency~$m\,c^2_0/h$
for a mass~$m$ \citep[cf.][]{Bro23}, because \emph{two} virtual gravitons have
to be emitted for one interaction, whereas in Paper~2 it is assumed that a
photon causes a reflection with an interaction rate of $\nu = E_\nu/h$ with
Planck's constant~$h$. The momentum transfer to a photon will thus be twice
as high as to a massive body with a mass equivalent to $E_\nu/c^2_0$.

If we apply the momentum conservation principle to photon-graviton pairs in
the same way as to photons \citep[cf.][]{LanLif}, we can write after a
reflection of $\vec {p}_{\rm G}$
%
\begin{eqnarray}
\vec {p}_\nu + 2\,\vec {p}_{\rm G} =
\vec {p}^*_\nu
\label{Eq:momentum}
\end{eqnarray}
with $|{\vec p}_{\rm G}| = p_{\rm G} = T_{\rm G}/c_0$.

We assume, applying Eq.~(\ref{Eq:momentum}) with
$p_{\rm G} \ll p_\nu = |\vec {p}_\nu|$, that
under the influence of a gravitational centre relevant
interactions occur on opposite sides of a photon with $p_{\rm G}$ and
$p_{\rm G}\,(1 - Y)$ transferring a net momentum of $2\,Y\,p_{\rm G}$. Note,
in this context, that the Doppler effect can only operate for interactions of
photons with massive bodies \citep[cf.][]{Fer32,Som78}.
Consequently, there will be no energy change of the photon, because both
gravitons are reflected with constant energies under these
conditions, and we can write for a pair of interactions
%
\begin{eqnarray}
E_\nu = |\vec p_\nu|\,c = |\vec p_\nu + 2\,Y\vec p_{\rm G}|\,c' =
|\vec p'_\nu|\,c' = E'_\nu ~,
\label{Eq:photon_energy}
\end{eqnarray}
where $\vec p'_\nu$ is the photon momentum after the events. If $\vec p_\nu$
and a component of $2\,Y\vec p_{\rm G}$ are pointing in the same direction,
it is $c' < c$, the speed is reduced; an antiparallel direction leads to
$c' > c$. Note that this could, however, not result in $c' > c_0$, because
$c = c_0$ can only be attained in a region with
an isotropic distribution of gravitons with a momentum of $p_{\rm G}$,
i.e. with a gravitational potential~$U_0 = 0$.

The momentum $\vec p_\nu$ of a photon radially approaching a gravitational
centre will be treated in line with Eq.~(6) in
Sect.~2 of Paper~2 for massive bodies, however, with twice the interaction
rate (valid for photons as explained above).
Since we know from observations that the deflection
of light during a close fly-by at the Sun is very small\,--\,to simplify the
calculations, we only treat this configuration\,--\,the
momentum variation caused by the weak and static
gravitational interaction is also very small.
The momentum change rate of the photon can then be approximated by
%
\begin{eqnarray}
\frac{\Del {\vec p}_\nu}{\Del t_M} \approx
2\,G_{\rm N}\,M\,\frac{\uvec r}{r^2}\,\frac{p_\nu}{c_0} ~,
\label{Eq:transfer_rate}
\end{eqnarray}
where $M$ the mass of the gravitational centre, $r = |\vec r|$ the
distance of the photon from the centre, and the position vector of
the photon is $r\,\uvec{r}$ with a  unit vector $\uvec{r}$.
The small deflection angle also allows us to approximate
the actual path by a straight line and use $x \approx c_0\,t_M$
along an $x$~axis.
The normalized momentum variation along the trajectory then is
%
\begin{eqnarray}
-\frac{c_0}{p_\nu}\left(\frac{\Del \vec{p}_\nu}{\Del t_M}\right)_x =
\frac{c_0}{p_\nu}\,\frac{\Del p_\nu}{\Del t_M}\cos \vartheta
\approx 2\,G_{\rm N}\,M\,\frac{x}{r^3}~.
\label{Eq:x_component}
\end{eqnarray}
The corresponding component perpendicular to the trajectory is
%
\begin{eqnarray}
-\frac{c_0}{p_\nu}\left(\frac{\Del \vec{p}_\nu}{\Del t_M}\right)_y =
\frac{c_0}{p_\nu}\,\frac{\Del p_\nu}{\Del t_M}\,\sin \vartheta
\approx 2\,G_{\rm N}\,M\frac{R}{r^3}~,
\label{Eq:y_component}
\end{eqnarray}
where $R$ is the impact parameter of the trajectory.
Integration of Eq.~(\ref{Eq:x_component}) over $t_M$ from $-\infty$ to $x/c_0$
yields
%
\begin{eqnarray}
\frac{1}{p_\nu}\,[\rmd \vec{p}_\nu(r)]_x \approx
\frac{2\,G_{\rm N}\,M}{c_0^2\,r} =
\frac{2\,G_{\rm N}\,M}{c_0^2\,\sqrt{R^2 + x^2}}~.
\label{Eq:x_integrated}
\end{eqnarray}

If we apply Eq.~(\ref{Eq:photon_energy}) to a photon approaching the
mass~$M$ along the $x$~axis
starting from infinity with $E_\nu = p_\nu\,c_0$, and
considering that the $y$~component in Eq.~(\ref{Eq:x_component}) is much smaller
than the x component in Eq.~(\ref{Eq:y_component}) for $x \gg R$,
the photon speed~$c(r)$ as a function of $r$
can be determined from
%
\begin{eqnarray}
p_\nu\,c_0 \approx \{p_\nu + [\rmd \vec{p}_\nu(r)]_x\}\,c(r)~.
\label{Eq:reduced_c}
\end{eqnarray}
Division by $p_\nu\,c_0$ then gives with Eq.~(\ref{Eq:x_integrated})
%
\begin{eqnarray}
\frac{1}{[n_{\rm G}(r)]_x} = \frac{c(r)}{c_0} \approx
1 - \frac{2\,G_{\rm N}\,M}{c_0^2~r} = 1 + \frac{2\,U(r)}{c_0^2}
\label{Eq:refraction}
\end{eqnarray}
as a good approximation of the inverse gravitational index of refraction
along the $x$~axis. The same index has been obtained albeit with different
arguments, e.g., by \citet{Booetal,YeLin}. The resulting speed of light is in
agreement with evaluations by \citet{Sch60}, for a radial
propagation\footnote{\citet{Ein12} states
explicitly that the speed at a certain location is not dependent on the
direction of the propagation.} in a central gravitational field, and
\citet{Oku00}\,---\,calculated on the basis of the standard Schwarzschild
metric. A decrease of the speed of light near the Sun, consistent with
Eq.~(\ref{Eq:refraction}), is not only supported by the predicted and
subsequently observed Shapiro delay
\citep{Sha64,Reaetal,Sha71,Kraetal,Baletal,KutZaj},
but also indirectly by the deflection of light \citep{Dysetal}.

\section{Gravitational redshift}
\label{sec:grav_red}

Since Einstein discussed the gravitational redshift and published conflicting
statements regarding this effect, the confusion could still not be cleared up
consistently \citep[cf., e.g.,][]{Man06,Sotetal}. In most of his publications
Einstein defined clocks as atomic clocks. Initially he assumed that the
oscillation of an atom corresponding to a spectral line might be an
intra-atomic process, the frequency of which would be determined by the atom
alone \citep{Ein08,Ein11}. \citet{Sco15} also felt that the equivalence
principle and the notion of an ideal clock running independently of
acceleration suggest that such clocks are unaffected by gravity. \citet{Ein16}
later concluded that clocks would slow down near gravitational centres thus
causing a redshift.

The question whether the gravitational redshift is caused by the emission
process (Case~a) or during the transmission phase (Case~b) is nevertheless
still a matter of recent debates. Proponents of (a) are, e.g.,
\citet{Moe57,Des90,Sch60,Craetal,Oha76,EarGly,Oku00,Okuetal} and of
(b): \citet{Hayetal,Feyetal,Str04,Fli06,Ran06,Wil06}.

There is general agreement on the observational and experimental facts and
most of the arguments are formally consistent with them, but different
physical processes or mathematical concepts are considered. In particular, it
is surprising that the same team of experimenters, albeit with different first
authors (Cranshaw et al. and Hay et al.) published different views on the
process of the Pound--Rebka--Experiment. \citet{PouSni} and \citet{Pou00}
pointed out, however,
that this experiment could not distinguish between the two options, because
the invariance of the velocity of the radiation had not been demonstrated.
\citet{Bon86} and \citet{Dic60} also left the question open. In many cases,
the confusion results from the unclear definitions of clocks and times as
detailed, for instance, by \citet{AshAll} and \citet{Oku00}.

\citet{Ein17} emphasized that for an elementary emission process not only
the energy exchange, but also the momentum transfer is of importance
\citep[cf., as well][]{Poi00,Abr03,Fer32}. Taking these considerations into
account, \citet{WilDwi14} formulated a photon emission process at
a gravitational potential~$U$ assuming that:
\begin{enumerate}
\item [(1)] The atom cannot sense the potential~$U$, in line with the
original proposal by \citet{Ein08,Ein11}, and initially emits the same
energy~$\Del E_0$ at $U > 0$ and $U_0 = 0$.
\item [(2)] It also cannot directly sense the speed of light at the location with a
potential~$U$. The initial momentum thus is $p_0 = \Del E_0/c_0$.
\item [(3)] As the local speed of light is, however, $c(U) \ne c_0$, a photon
having an energy of $\Del E_0$ and a momentum $p_0$ is not able to propagate.
The necessary adjustments of the photon energy and momentum as well as the
corresponding atomic quantities then lead in the \emph{interaction region} to
a redshift consistent with $h \nu = \Del E_0\,(1 + U/c^2_0)$ and observations.
\end{enumerate}

As outlined in Sect.~\ref{sec:graviton}, there is general
agreement in the literature that the local speed of light is
%
\begin{eqnarray}
c(U) \approx c_0 \left(1 + \frac{2\,U}{c^2_0}\right)
\label{Eq:local_speed}
\end{eqnarray}
in line with Eq.~(\ref{Eq:refraction}). It has, however, to be noted that
in Sect.~\ref{sec:graviton} the speed~$c(U)$ was obtained for a photon
propagating from $U_0$ to $U$, and, therefore, the physical process which
controls the speed of newly emitted photons is not established. An attempt
to do that will be made in the next section.

\section{An aether model with photons as solitons}
\label{sec:aether}
Before we suggest a specific aether model, a few statements on the aether
concept in general should be mentioned. Following \citet{MicMor} famous
experiment, \citet{Ein05a,Ein08} concluded that the concept of a light aether
as carrier of the electric and magnetic forces is not consistent with the STR.
In response to critical remarks by \citet{Wie11}, cf. \citet{Sch90} for
Wiechert's support of the aether, \citet{Lau12} wrote that the existence
of an aether is not a physical, but a philosophical problem, but later
differentiated between the physical world and its mathematical formulation.
A four-dimensional `world' is only a valuable mathematical trick; deeper
insight, which some people want to see behind it, is not involved \citep{Lau59}.

In contrast to his earlier statements, Einstein said at the end of a speech
in Leiden that according to the GTR a space without aether cannot be
conceived \citep{Ein20}; and even more detailed: Thus one could instead of
talking about `aether' as well discuss the `physical properties of space'.
In theoretical physics we cannot do without aether, i.e., a continuum endowed
with physical properties \citep{Ein24}.
\citet{Mic28} confessed at a meeting in Pa\-sa\-de\-na in the
presence of H.A. Lorentz that he clings a liitle to the aether; and
\citet{Dir51} wrote in a letter to Nature that there are good reasons for
postulating an {\ae}ther.

\citet{WiDwWi} proposed an impact model for the electrostatic force
based on massless \emph{dipoles}.
The vacuum is thought to be permeated by these dipoles that are, in the absence of
electromagnetic or gravitational disturbances, oriented and directed randomly
propagating along their dipole axis with a speed of~$c_0$. There is little or
no interaction among them. Note that such electric dipoles have no mean
interaction energy, even in the classical theory \citep[see, e.g.,][]{Jac06}.
We suggest to identify the dipole distribution with an aether. This is very
similar to the conclusion of \citet{Pre75}:
\begin{enumerate} \item[ ]
``[...] first, that the normal
state of the component particles of the ether is a state of motion; second,
that this motion of the particles takes place in straight lines; and third,
that this motion takes place towards every possible direction.''
\end{enumerate}
Einstein's aether mentioned above may, however, be more related to the
gravitational interactions \citep[cf.][]{Gra01}. In this case, we have to
consider the graviton distribution as another component of the aether.

If we assume that an individual dipole interacts with gravitons in the same
way as photons, see Eq.~(\ref{Eq:photon_energy}), according to
%
\begin{eqnarray}
E_{\rm D} = |\vec p_{\rm D}|\,c = |\vec p_{\rm D} + 2\,Y\vec p_{\rm G}|\,c' =
|\vec p'_{\rm D}|\,c' = E'_{\rm D} ~,
\label{Eq:dipole_energy}
\end{eqnarray}
where $E_{\rm D}$ and $\vec p_{\rm D}$ refer to the energy and momentum of a
dipole. We can then modify Eqs.~(\ref{Eq:transfer_rate}) to (\ref{Eq:reduced_c})
by changing $\nu$ to D and find that Eqs.~(\ref{Eq:refraction}) and
(\ref{Eq:local_speed} ) are also valid for dipoles with a speed of $c_0$ for
$U_0 = 0$. One exception from Preston's ``ether'' is that dipoles can, according
to a modified Eq.~(\ref{Eq:y_component}), be deflected by graviton
interactions.

Considering that many suggestions have been made to describe photons as
solitons \citep[e.g.,][]{Dir27,Vig91,KamSla,Meu13,Ber13,Beretal}, we
also propose that a photon is a soliton propagating in the dipole
aether with a speed of~$c(U)$, cf., Eq.~(\ref{Eq:local_speed}), controlled by
the dipoles moving in the direction of propagation of the photon.
The dipole distribution thus determines the gravitational index of
refraction, cf. Eq.~(\ref{Eq:refraction}), and consequently the speed of
light~$c(U)$ at the potential~$U$. This solves the problem formulated
at the end of Sect.~\ref{sec:grav_red} and might be relevant for other
phenomena, such as gravitational lensing and the cosmological redshift
\citep[cf., e.g.,][]{Ell10,CheKan}.

We will further assume that the dipoles constituting a photon will have turned
the orientation of their axes to a direction perpendicular to the photon
velocity vector. This avoids any electrostatic interactions
during emission and absorption processes of photons, and will probably also be
required by their polarization effects.

\section{Discussion and Conclusion}
\label{sec:concl}
Our aim was to identify a physical process that leads to a speed~$c(U)$ of
photons controlled by the gravitational potential~$U$. This could be achieved
by postulating an aether model with moving dipoles, in which a
gravitational index of refraction $n_{\rm G}(U) = c_0/c(U)$ regulates the
emission and propagation of photons as required by energy and momentum
conservation principles. The emission process thus follows Steps~(1) to (3)
in Sect.~\ref{sec:grav_red}, where the local speed of light is given by the
gravitational index of refraction~$n$. In this sense, the statement that an
atom cannot detect the potential~$U$ by \citet{Mueetal} is correct; the local
gravity field~$\vec{g}$, however, is not controlling the emission process.

A photon will be emitted by an atom with appropriate energy and momentum
values, because the local speed of light requires an adjustment of the
momentum. This occurs in the interaction region between the atom and its
environment as outlined in Step~(3) of Sect.~\ref{sec:grav_red}. A receiver
of the same type next to the emitter would also not be able to determine the
potential either, because the energy and momentum restrictions apply for the
absorption process as well.


\acknowledgments
This research has made extensive use of the
Smithsonian Astrophysical Observatory (SAO)/National Aeronautics and Space
Administration (NASA) Astrophysics Data System (ADS).
Administrative support has been provided by the Max-Planck-Institute for Solar
System Research and the Indian Institute of Technology (Banaras Hindu
University).

\end{document}